\documentclass[12pt,a4paper]{article}

\usepackage[]{graphicx}
\usepackage{amsmath,amssymb}
\usepackage{cite}
\usepackage{subfigure}
\usepackage{color}

\def\BibTeX{{\rm B\kern-.05em{\sc i\kern-.025em b}\kern-.08em
		T\kern-.1667em\lower.7ex\hbox{E}\kern-.125emX}}
\usepackage{booktabs}
\usepackage{multirow}
\usepackage{array}
\newcolumntype{+}{>{\global\let\currentrowstyle\relax}}
\newcolumntype{^}{>{\currentrowstyle}}

\newcommand{\VTH}{\ensuremath{V_{th}}}

\newcommand{\VDS}{\ensuremath{\textrm{V}_{DS}}}

\newcommand{\VGS}{\ensuremath{\textrm{V}_{GS}}}

\newcommand{\IDS}{\ensuremath{\textrm{I}_{DS}}}

\begin{document}
	\title{Multi-level Operation of FeFETs Memristors:\\ the Crucial Role of Three Dimensional Effects}
	\author{\small{Daniel Lizzit, Thomas Bernardi and David Esseni}\\
		\small{DPIA, University of Udine,
		via delle Scienze 206, 33100 Udine, Italy}\\
		\small{e-mail: daniel.lizzit@uniud.it}\\
	}
	
	\date{\vspace{-5ex}}
	\maketitle
	\begin{abstract}
		
		This paper investigates and compares through a comprehensive TCAD analysis 2D and 3D simulations for ferroelectric based FETs.
		We provide clear evidence that the multiple read conductance values experimentally observed in FeFETs stem from source to drain percolation current paths, which are governed by the polarization patterns in the ferroelectric domains.
		Such a physical picture makes 3D simulations indispensable to capture even the qualitative features of the device behaviour, not to mention the quantitative aspects.
		
	\end{abstract}
 \textbf{© 2023 IEEE. Personal use of this material is permitted. Permission from IEEE must be obtained for all other uses, in any current or future media, including reprinting/republishing this material for advertising or promotional purposes, creating new collective works, for resale or redistribution to servers or lists, or reuse of any copyrighted component of this work in other works.\\
	\\
https://ieeexplore.ieee.org/document/9947169}

	\section{Introduction}
	\label{Sec:Intro}
	%
	Ferroelectric based FETs (FeFETs) are very promising memories and  memristors for neuromorphic computing \cite{Slesazeck_Nanotech2019,Covi_NeuroCompAndEngineering2022}. 
	Unlike other device concepts, such as resistive RAMs (ReRAM) or Phase Change Memories (PCM) \cite{Zahoor_NanoscaleResLett2020}, that require a current flow for the switching operation,  the physical substrate offered by FeFETs exploits a displacement current to set the ferroelectric polarization, thus holding the promise of low-power 
	consumption memristors \cite{Majumdar_Nanotech2019}.
	
	Ferroelectric oxides for CMOS compabile devices are based either on hafnium oxides with Si doping (HSO) or on hafnium zirconium oxides (HZO), both exhibiting  remnant polarization values ranging from 15 to 30 $\mu$C/cm$^2$ and coercive fields in the range of 1-2 MV/cm \cite{Trentzsch_IEDM2016,Jerry_IEDM2017,Kim_Nanoscale2020, Zhao_TED2022}. 
	
	It is well known that ferroelectric materials tend to minimize their free energy by creating domains \cite{Park_AdvMater2019,Park_AdvFunMater2022}, whose size is usually associated to the size of the polycrystalline grains \cite{Mulaosmanovic_ACSAMI2017}. 
	Indeed, for the film thicknesses of practical interest for nanoscaled electronic devices, namely in the range of 5-10 nm, ferroelectric oxides are polycristalline with a grain size in the nanometer range \cite{Mulaosmanovic_ACSAMI2017,Lee_APR2021}. 
	
	In this paper, a physical based TCAD simulation approach is used to account for the multi-domain nature of the ferroelectric material and thus explore a multi-level operation of FeFETs memristors.
	
	\section{Modelling of the FeFET}
	
	Simulations are carried out by using the TCAD tool Sentaurus-Device package \cite{SdeviceManual}, that couples the drift-diffusion equations with a kinetic model for the ferroelectric dynamics based on the phenomenological  Landau-Ginzburg-Devonshire (LGD) equation \cite{SdeviceManual}.  
	
\begin{figure}[h!]
		\centering
		\includegraphics[width=1.00\hsize]{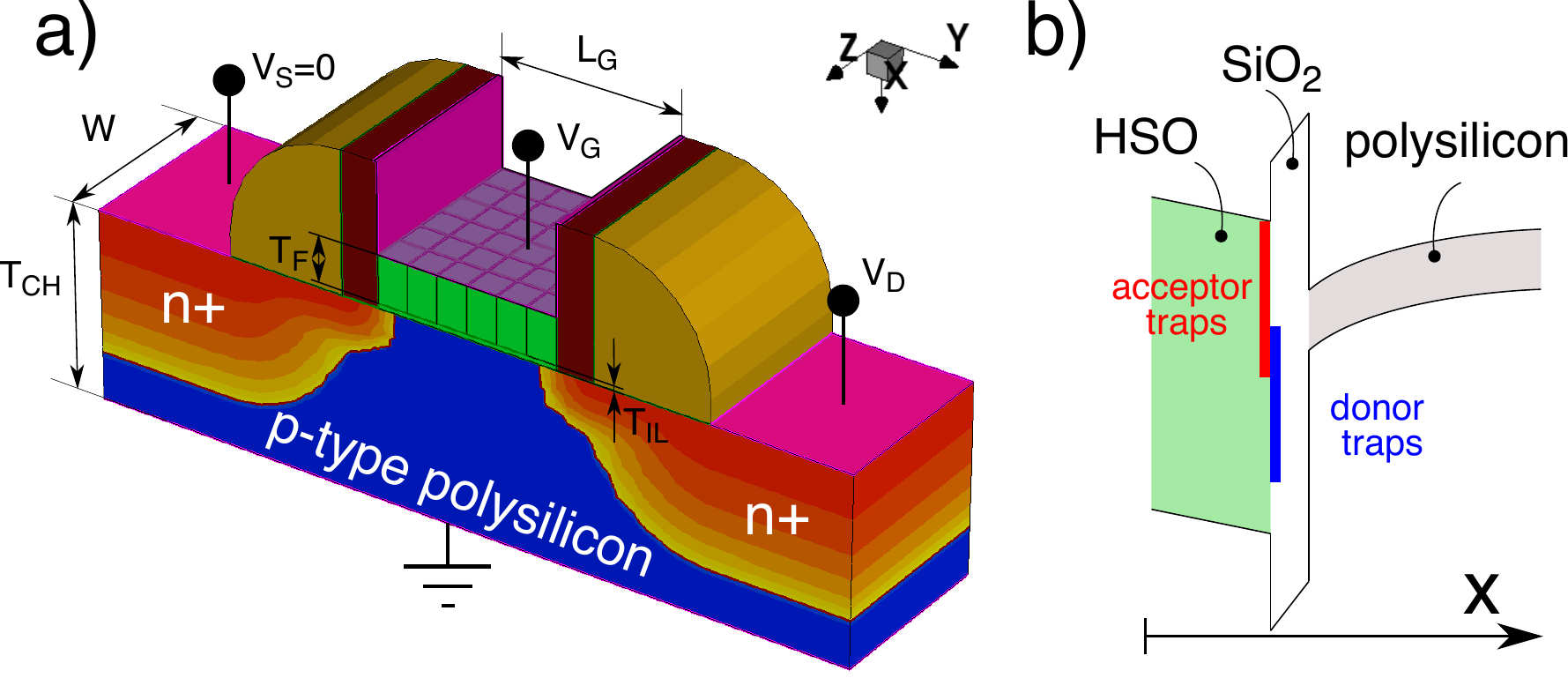}
		\vspace{-4mm} \caption{
			a) Simulated FeFET with a gate length $L_G$$=$40 nm and a gate width $W$$=$40 nm, and consisting of 6$\times$6 domains of ferroelectric HSO with a 10 nm thickness. 
			The dielectric spacer between each grain is 5 \AA\ thick. The interfacial SiO$_2$ layer has a thickness $t_{IL}$$=$1 nm. 
			Source and drain regions are $n$-type doped with a density of 5$\times$10$^{20}$ cm$^{-3}$, wherease the channel is $p$-type with a doping of 1$\times$10$^{18}$ cm$^{-3}$ and a thickness $t_{CH}$$=$ 40 nm.
			b) Sketch of the bands profile in the device gate stack and of the acceptor and donor type traps at the FE/IL interface, having a uniform density of respectively  D$_{it;acc}$=0.8$\times$10$^{14}$ cm$^{-2}$eV$^{-1}$ and D$_{it;don}$=1.0$\times$10$^{14}$cm$^{-2}$eV$^{-1}$.}
		\label{Fig:DeviceSketch}
	\end{figure}

	A sketch of the simulated device is shown in Fig.\ref{Fig:DeviceSketch}, where the polycristalline nature of the ferroelectric HfO$_2$ is described by defining square ferroelectric domains having a 6 nm size in the $y$-$z$ plane, namely in the range of experimentally reported values \cite{Mulaosmanovic_ACSAMI2017,Lee_APR2021}. An essentially  uniform polarization inside each grain is enforced by using an artificially large value of  the domain wall coupling parameter \cite{SdeviceManual}. 
	Ferroelectric domains are separated by a thin non-ferroelectric spacer having the same dielectric constant as HSO ({\it i.e.} $\varepsilon_{FE}$=30 \cite{Muller_ECS2015}).
	The electrostatic coupling between the domains is inherently accounted for in simulations. However the model does not assign any energy penalty to the formation of domains with an anti-parallel polarization or, equivalently, to the formation of 180° domain walls. This is consistent with	recent {\it ab-initio} calculations of the domain wall energy in HfO$_2$ based ferroelectrics \cite{Lee_Science2019}.

	The calibration of the Landau's anysotropic constants has been previously carried out by comparison  against experimental data for an HSO capacitor  \cite{Lizzit_ESSDERC2021}, leading to $\alpha$=$-5.37$$\times 10^8$ m/F, $\beta$=$9.62$$\times 10^8$ m$^5$/(FC$^2$), $\gamma$=$9.59$$\times 10^{10}$ m$^9$/(FC$^4$),  corresponding to a remnant polarization P$_R$$=$20 $\mu$C$/$cm$^2$ and a coercive field E$_C$$=$1.1 MV$/$cm.
	Our simulations include a domain to domain statistical dispersion of the ferroelectric properties by using a normal distribution for the coercive electric field featuring a  mean value E$_C$$=$1.1 MV$/$cm  and a standard deviation $\sigma_{E_C}$=0.3E$_{C,0}$. The resistivity for
the ferroelectric switching employed in simulations is $\rho$=110
$\Omega$m \cite{Kobayashi_IEDM2016},  resulting in a time constant $\tau$=$\rho$/2$|\alpha|$$\sim$100 ns.
	
	In this work we consider a channel length of 40 nm, corresponding to six  ferroelectric domains in the source to drain direction.
	Furthermore, we denote as 2D  simulations those including a single domain in the device width direction (i.e. 6$\times$1 FeFET), and as 3D simulations those featuring more domains in the transverse direction (e.g. 6$\times$3 and 6$\times$6 FeFETs).
	Between the ferroelectric oxide and the polysilicon channel, which is compatible with a BEOL integration, the simulated FeFET has a 1 nm thick SiO$_2$ interlayer (IL), which is typically observed in conventional growth processes for hafnia-based ferroelectrics on silicon \cite{Mulaosmanovic_TED2021, Trentzsch_IEDM2016}.  
	This layer plays an important role in the modelling of the device  in several respects. 
	Indeed, experimental results based on Hall and split-CV measurements suggest that most of the charge compensating for the ferroelectric polarization is located in the IL at the ferroelectric-dielectric interface \cite{Toprasertpong_IEDM2019}, as  it's been also confirmed in more recent experiments and simulations \cite{Zhao_TED2022}. 
	Moreover, in the absence of traps, the voltage drop across the IL reduces the drop on the ferroeletric and can lead to the so called minor-loop operation, where domains with trajectories located along nonsaturated hysteresis loops tend to back-switch after the write programming phase. 
	Indeed, we have previously shown that, by neglecting altogether the charge trapping, simulations exhibit irreconcilable discrepancies  with the experimental results \cite{Lizzit_ESSDERC2021}.
	Based on these premises, in the simulations of this work we employed an interfacial trap density at the FE-IL interface as reported in Fig. \ref{Fig:DeviceSketch}b.
	Interfacial traps in this work can exchange carriers with the polysilicon channel through a non-local trapping model that includes both elastic and phonon-mediated tunneling processses across the SiO$_2$ IL \cite{SdeviceManual}. 
	Tunneling emission and capture rates are described by the trap volume, the Huang-Rhys factor, the phonon energy and the tunneling masses \cite{SdeviceManual}, which can be regarded as fitting parameters and for the SiO$_2$ IL are taken from Ref. \cite{Vandelli_TED2011}.

	Simulations are performed by using a constant carrier mobility of 10 cm$^2$/Vs in the polysilicon  channel \cite{Lifshitz_TED1994}. The \VDS\ in this work is 50 mV, therefore the longitudinal electric fields  are such that velocity saturation effects are negligible. 
	
	In these simulations, the bottom of the polysilicon channel is  connected to ground to effectively provide majority carriers (holes)  that allow for a positive to negative polarization switching within tens or hundred of nanoseconds. 
 	This simulation setup can still represent a BEOL compatible FeFET with the channel lying on a thick oxide layer and that is electrically contacted from the top side.

	%
	\section{Simulation results}
	\label{Sec:2}
	
\begin{figure}[h!]
		\centering
		\includegraphics[width=1.00\hsize]{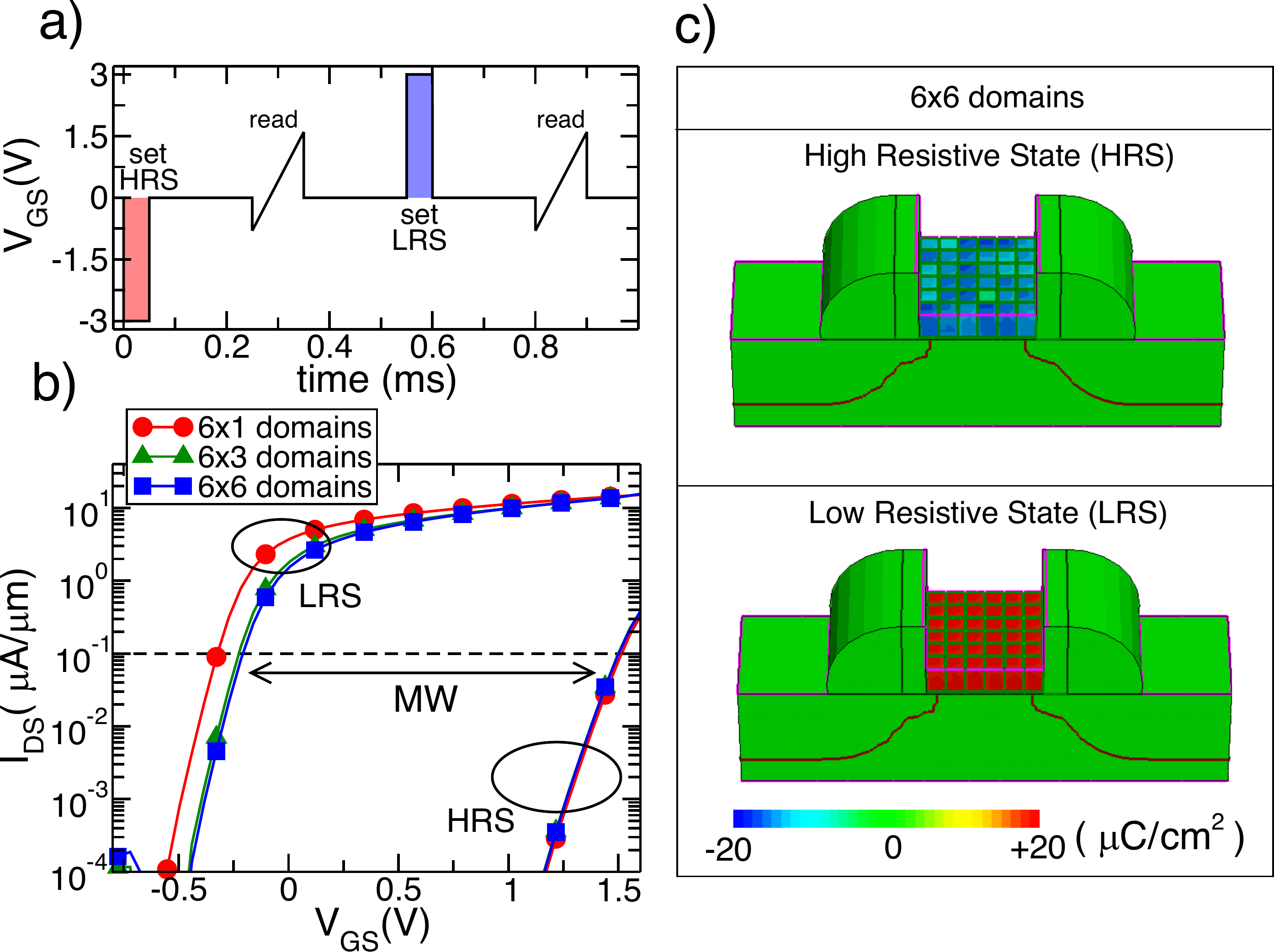}
		\vspace{-4mm} \caption{
			a) Pulsed waveforms for the write operation have duration t$_P$=50 $\mu$s and are followed by a gate sweep from -1 to 1.6V to perform the readout. b) \IDS\ vs. \VGS\ curves at \VDS = 0.05 V for the HRS and LRS in a FeFET with 6 domains along the transport direction and 1, 3 or 6 domains along the device width. The horizontal dashed line corresponding to 100 nA/$\mu$m is used to determine the MW width. c) Domain polarization patterns for the 6x6 domains FeFET obtained at the maximum read voltage amplitude \VGS =1.6V.}
		\label{Fig:Figure_2}
	\end{figure}

	To investigate the maximum shift of the threshold voltage (\VTH ), or equivalently  the memory window (MW)  of simulated FeFETs, we first used a positive and negative gate voltage with sufficient amplitude and duration to drive the device in the low resistance (LRS) and high resistance state (HRS), where all the domains have respectively positive or negative polarization\footnote{The polarization is defined positive when it points to the transistor channel, namely along the $x$ direction in Fig.\ref{Fig:DeviceSketch}.}.
	Figure \ref{Fig:Figure_2}a shows the applied gate voltage scheme, consisting of a write pulse followed by a \VGS\ ramp for the readout (with a read voltage small enough to leave the polarization pattern unperturbed). The \IDS\ versus \VGS\  characteristics are shown in Fig. \ref{Fig:Figure_2}b  for a 6x1, a 6x3 and a  6x6 domains FeFET. In the three devices  the MW is essentially the same and equal to approximately 1.7 V. Hence, the uniform polarization state enforced by the write pulse makes the device behaviour insensitive to the number of domains in the width direction.
	Such a simulated MW is in good agreement with the experimental values for the HSO based FeFETs reported in Refs. \cite{Dunkel_IEDM2017,Mulaosmanovic_ACSAMI2017}.

\begin{figure}[h!]
		\centering
		\includegraphics[width=0.8\hsize]{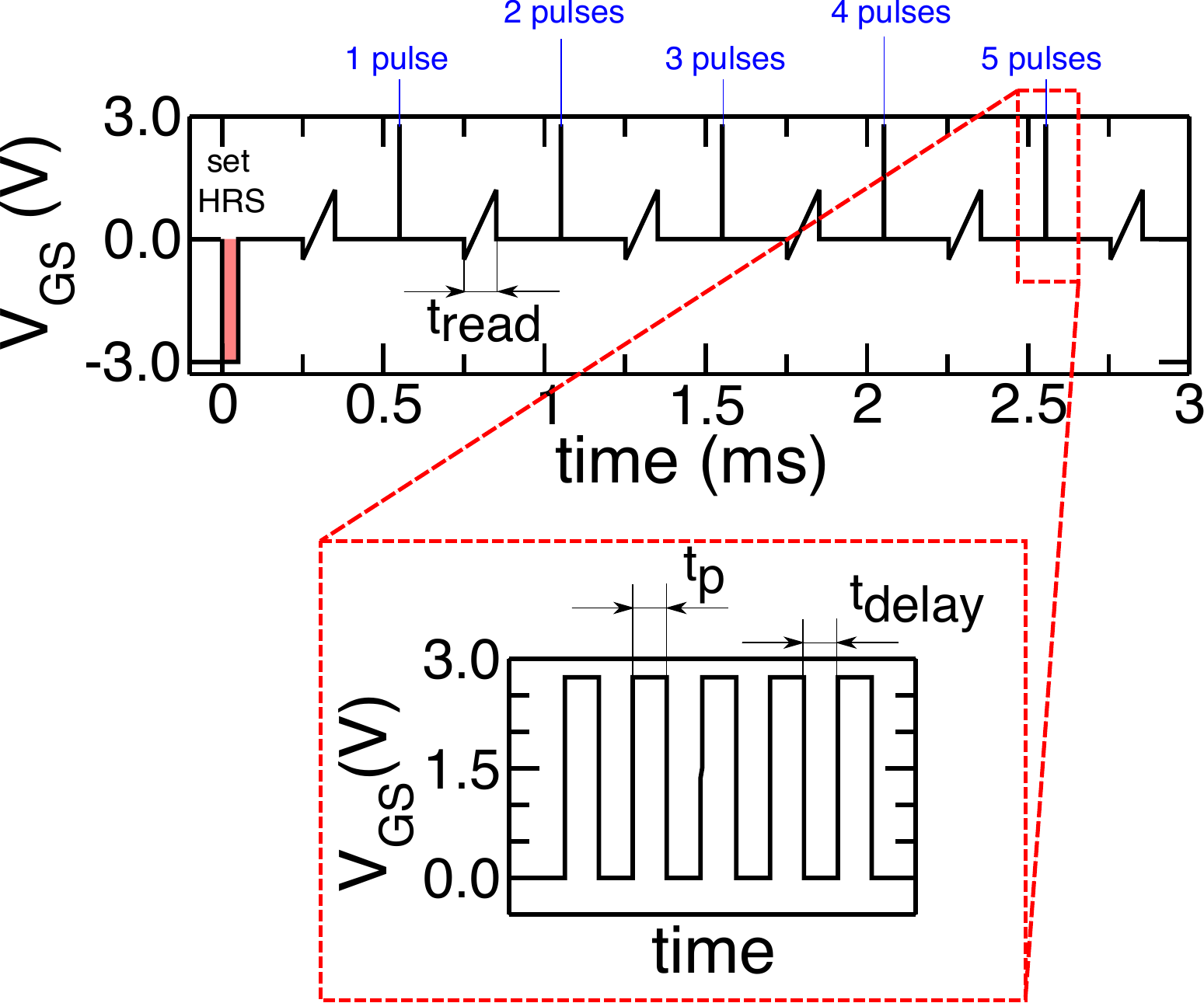}
		\vspace{-4mm} \caption{
			Pulse scheme for potentiation simulations. After a pre-set signal with amplitude of -3 V, a series of pulses with duration t$_p$=200 ns separated by t$_{delay}$=200 ns and with amplitude of 2.75 V are applied to the gate contact. Between each series of pulses a read operation is performed by sweeping the \VGS\  in the range  -0.5 to 1.2 V in 100 $\mu$s.}
			\label{Fig:V_time}
	\end{figure}

	In order to explore with simulations the possibility of setting multiple \VTH\ values, we applied a series of increasing number of constant amplitude pulses at the gate terminal followed by a read sweep, as illustrated by the waveforms in Fig. \ref{Fig:V_time}.
	The idea behind this pulse scheme is to exploit the accumulative switching behaviour, where multiple pulses separated by a time delay, t$_{delay}$, can induce a progressive polarization switching involving very few and possibly even a single domain \cite{Covi_NeuroCompAndEngineering2022,Jerry_IEDM2017}.

\begin{figure}[h!]
			\centering
			\includegraphics[width=1.00\hsize]{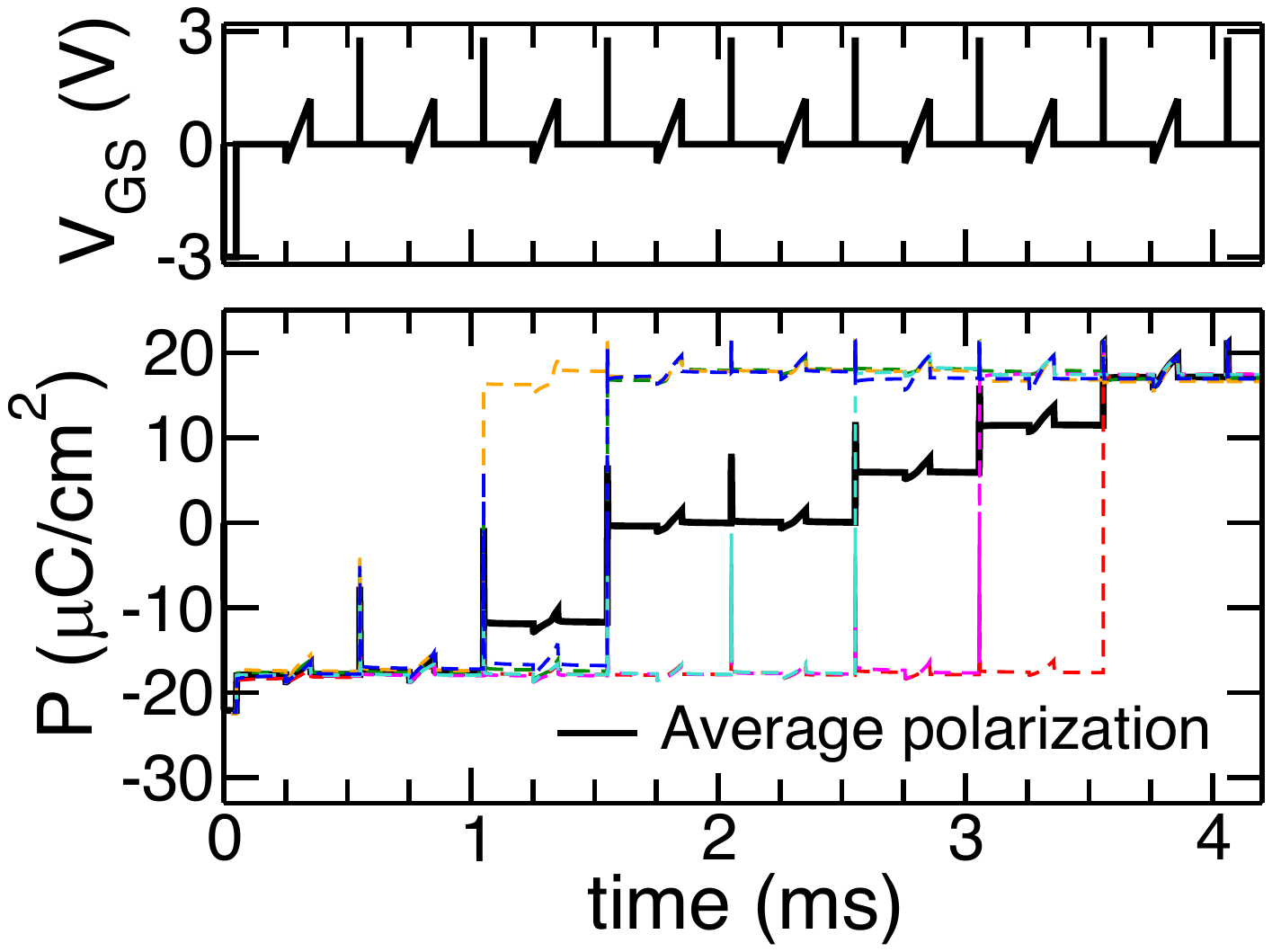}
			\vspace{-14mm} \caption{
				Top: \VGS\ waveforms with write pulses separated by a read sweep. Bottom: polarization trajectories for the 6$\times$1 FeFET of each domain (dashed line) and average value (solid line). All domains have positive polarization  after 28 pulses.}
			\label{Fig:6x1_Ptime}
		\end{figure}

	Figure \ref{Fig:6x1_Ptime} shows the polarization trajectories of each domain (dashed, coloured lines) for the 6$\times$1 FeFET. The negative to positive switching of polarization in one or a few domains occurs almost at each group of \VGS\ pulses, whereas the read operation does not perturb the stored polarization.
	
\begin{figure}[h!]
			\centering
			\includegraphics[width=0.9\hsize]{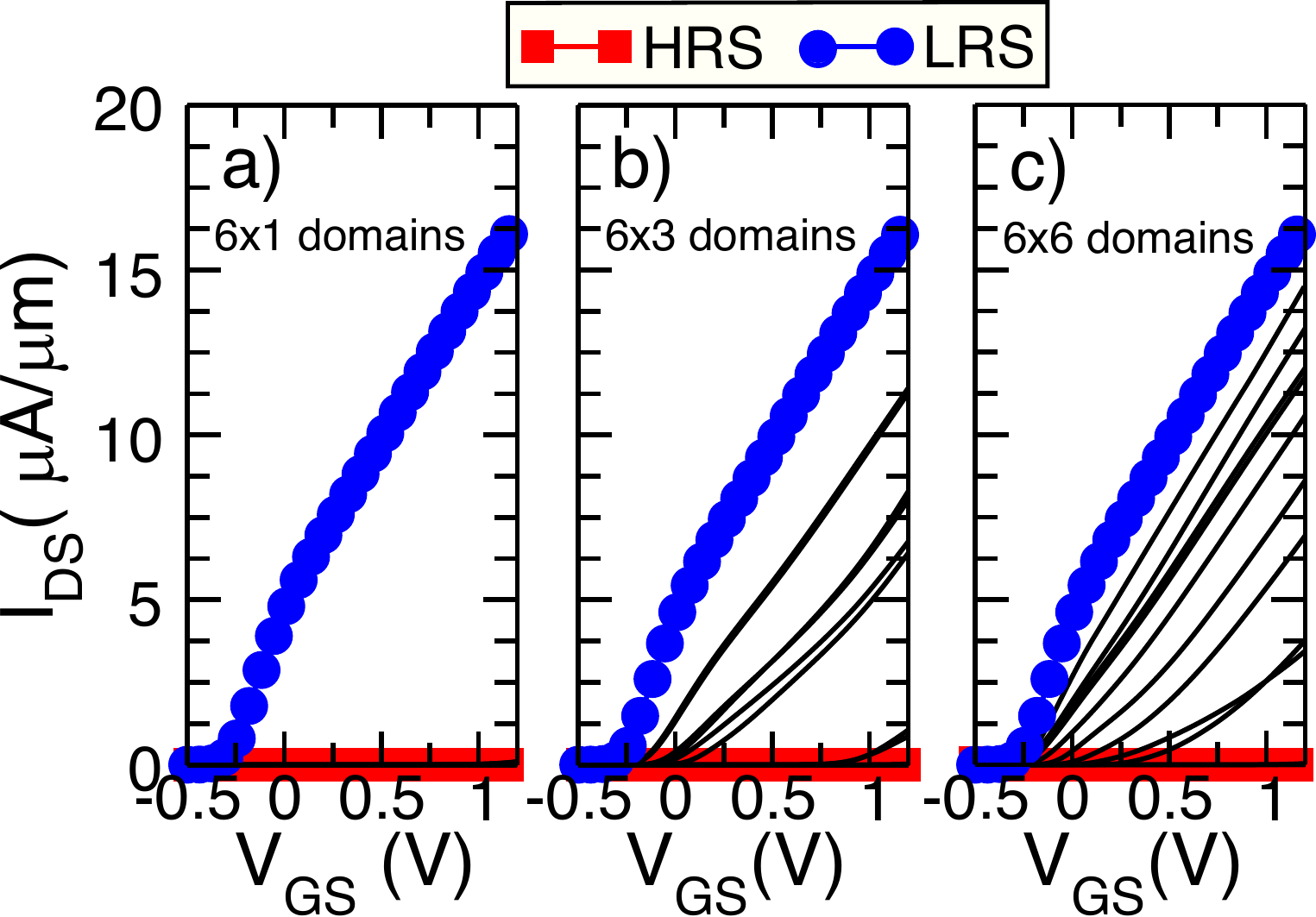}
			\vspace{-4mm} \caption{ \IDS - \VGS\ curves for the a) 6$\times$1, b) 6$\times$3 and c) 6$\times$6 domains FeFETs obtained after series of gate pulses with same amplitude and duration as shown in Fig. \ref{Fig:V_time}. Simulations corresponding to \VDS = 0.05 V.}
			\label{Fig:I_V_differentDomains}
		    \end{figure}

\begin{figure}[h!]
			\centering
			\includegraphics[width=0.80\hsize]{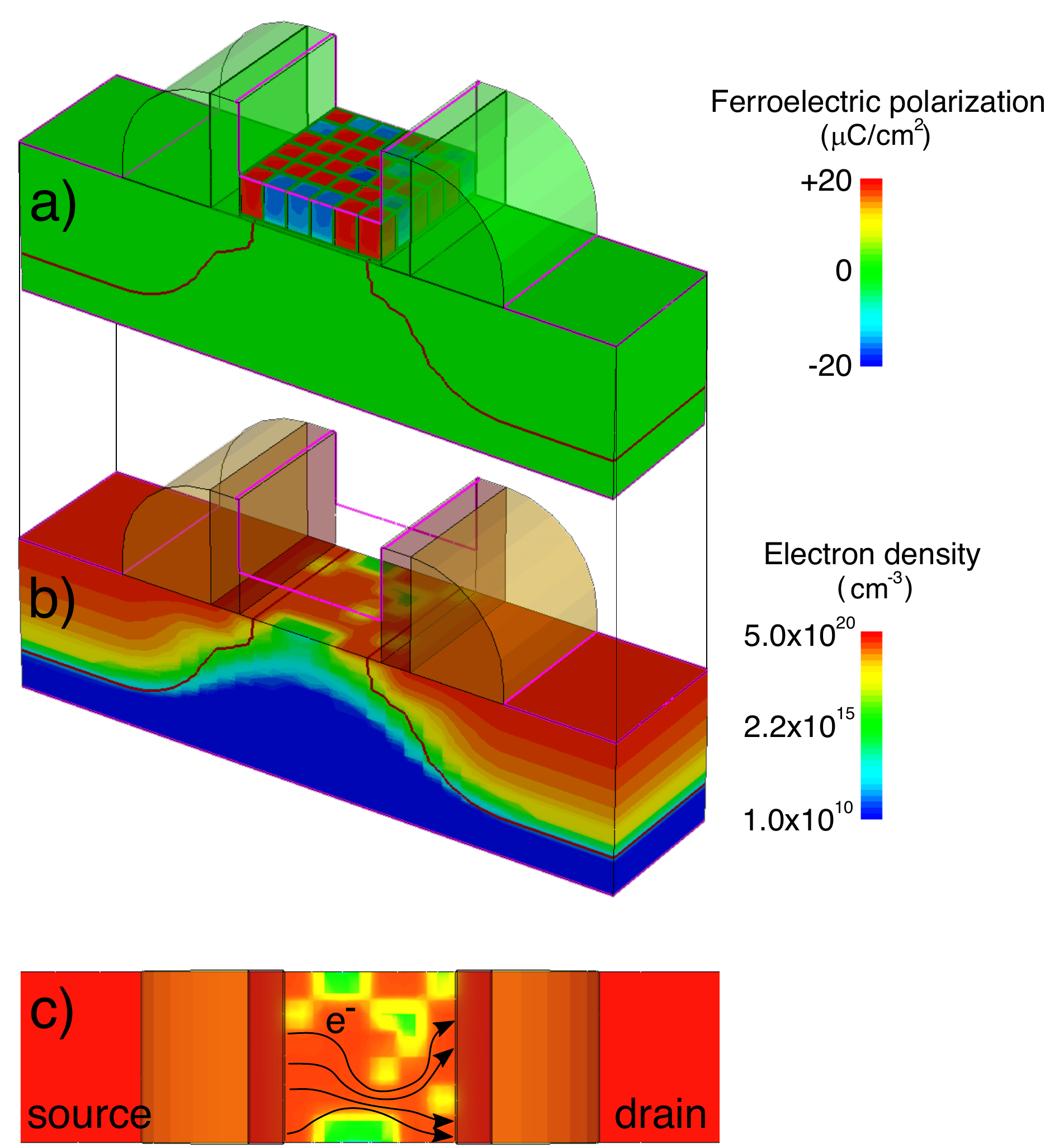}
			\vspace{-0mm} \caption{ Simulation results for the 6$\times$6 FeFET.
				a) ferroelectric polarization, b) electron density obtained for the maximum reading voltage of \VGS = 1.2 V. c) sketch of the percolation paths from source to drain.}
			\label{Fig:Plot_6x6_PolarizationAndCurrentDensity}
		\end{figure}

	Figure \ref{Fig:I_V_differentDomains}  shows the	\IDS\ versus \VGS\ curves obtained during the $V_{GS}$ ramp used for the readout operation (see Fig.\ref{Fig:V_time}). The FeFET with a single domain along the width direction ({\it i.e.} the 6$\times$1 FeFET shown in Fig. \ref{Fig:I_V_differentDomains}a) exhibits a digital behaviour. 
	This is because any domain with negative polarization precludes the formation of  a source to drain conduction path. Therefore the FeFET is basically in the LRS when all domains have a positive polarization, and it is in the HRS for all other polarization patterns.
	On the other hand, FeFETs having 3 and 6 domains along the device width exhibit several intermediate \VTH\ values or, equivalently, intermediate current levels in the readout mode.
	In order to illustrate the physics behind this multi level behaviour, Fig. \ref{Fig:Plot_6x6_PolarizationAndCurrentDensity}a shows the polarization pattern for the 6$\times$6 domains FeFET after 
	15 potentiation pulses. From the electron density map distribution in Fig. \ref{Fig:Plot_6x6_PolarizationAndCurrentDensity}b-c it can be observed the formation of high conductance paths from source to drain. The number of these conductance paths enlarges by increasing the number of potentiation pulses, which results in the multi-level operation observed in Fig. \ref{Fig:I_V_differentDomains}b-c.
		
\begin{figure}[h!]
			\centering
			\includegraphics[width=0.80\hsize]{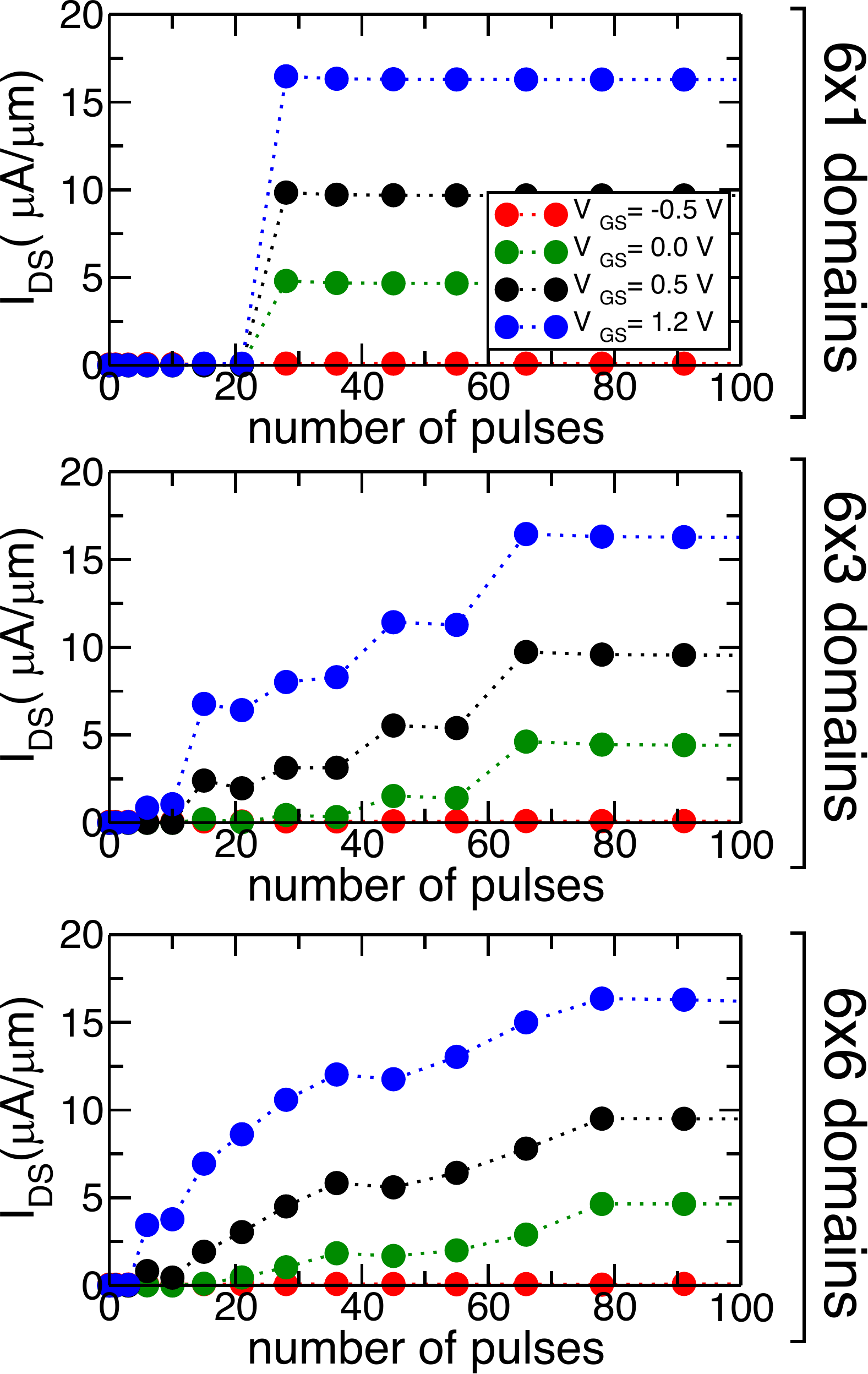}
			\vspace{-0mm} \caption{
				FeFET simulated channel current versus number of pulses for the 6$\times$1, 6$\times$3 and 6$\times$6 FeFETs, and for different \VGS\ values. 
				Results obtained for \VDS =0.05 V. 
				}
			\label{Fig:I_Pulses}
		\end{figure}

		The read current corresponding to different read $V_{GS}$ values is illustrated in Fig. \ref {Fig:I_Pulses} versus the number of potentiation pulses and for the different simulated structures. 
		Consistently with the results shown in Fig. \ref{Fig:I_V_differentDomains}, only a single LRS is observed for the 6$\times$1 FeFET for any read $V_{GS}$, which is reached after 28 potentiation pulses that set all the ferroelectric domains to a positive polarization. 
		The 6$\times$3 FeFET exhibits instead a multi-level behaviour, where discrete read $I_{DS}$ values are observed by increasing the number of pulses (and for a fixed read $V_{GS}$), corresponding to the activation of new percolation current paths or the reinforcement of existing ones. Hence the read current in the 6$\times$3 FeFET shows some discrete values which are modulated by the potentiation pulses. Finally, Fig. \ref {Fig:I_Pulses} shows that the 6$\times$6 FeFET offers an almost continuous set of read $I_{DS}$ values, in virtue of the much larger number of possible source to drain percolation current paths. Consistently with the results in Fig. \ref{Fig:I_V_differentDomains}, the three FeFETs have essentially the same read $I_{DS}$ per unit width in the LRS.
		
		
		\section{Discussion and conclusions}
        
        This work has presented a comprehensive TCAD study of the multi-level operation in FeFETs memristors, focusing in particular on the difference between 2D and 3D simulations. We found that, if an accumulative switching behavior can be induced in the HSO film, then the modulation of the percolation current paths in the device offers a powerful pathway for a multi-level memristor operation.
        
        Despite the encouraging results in Fig. \ref{Fig:I_Pulses}, it should be noticed that the displayed behaviour depends critically on a number of simulation parameters, including the density of traps and their emission and capture rates, as well as the features of the \VGS\ waveforms including the delay times between the groups of potentiation pulses. In our simulations, for instance, the accumulative switching behavior is linked to the trapping dynamics, and it is understood that the properties of traps cannot be regarded as a design knob.
        Nevertheless, trap densities can be partly engineered in ALD deposited Al$_2$O$_3$ films by tuning the deposition parameters, such as temperature, oxidant precursor type and dosing time \cite{Rahman_Materials2020}.
        
        Although the material and device engineering for FeFETs memristors is still a challenging path, our simulations predict good potentials for this class of ferroelectric devices.
        
		
		
		\vspace{3mm}
		\noindent {\bf Acknowledgments}
		This work was supported by the European Union through the BeFerroSynaptic project (GA:871737).

	\end{document}